\documentclass[12pt]{article}

\usepackage{amsthm,amsmath,amssymb}
\usepackage{graphicx}

\newcommand{\bgamma}{\boldsymbol{\gamma}}
\newcommand{\fl}{}%{\hspace*{-6pc}}
\newcommand{\case}[2]{{\textstyle\frac{#1}{#2}}}
\newcommand{\rmd}{\mathrm{d}}
\newcommand{\Or}{\mathord{\mathrm{O}}}

\newcommand{\C}{\mathbb{C}}
\newcommand{\Disk}{\mathbb{D}}
\newcommand{\Half}{\mathbb{H}}
\newcommand{\R}{\mathbb{R}}
\newcommand{\Prob}{\mathbf{P}}
\newcommand{\bd}{\partial}  
\newcommand{\w}{\mathbf{w}}

\newcommand{\z}{\mathbf{z}}
\newcommand{\hatgamma}{\hat{\bgamma}}
\newcommand{\intprob}{\Phi}
\newcommand{\ccharge}{\mathbf{c}}
\newcommand{\bargamma}{\bgamma}
\newtheorem{theorem}{Theorem}[section]
\newtheorem{proposition}[theorem]{Proposition}
\theoremstyle{remark}
\newtheorem*{remark}{Remark}

\title{Using the Schramm-Loewner evolution to explain certain non-local observables in the 2d critical Ising model}

\author{
Michael J.~Kozdron\footnote{Research supported in part by the Natural Sciences and Engineering Research Council of Canada.}\\
{\small Department of Mathematics \& Statistics, College West 307.14}\\
{\small University of Regina, Regina, SK S4S 0A2 Canada}\\ 
{\small \texttt{kozdron@stat.math.uregina.ca}}
}

\date{}

\begin{document}

\maketitle

\begin{abstract}
We present a mathematical proof of theoretical predictions made by Arguin and Saint-Aubin, as well as by Bauer, Bernard, and Kyt\"ol\"a, about certain non-local observables for the two-dimensional Ising model at criticality by combining Smirnov's recent proof of the fact that the scaling limit of critical Ising interfaces can be described by chordal SLE$_3$ with Kozdron and Lawler's configurational measure on mutually avoiding chordal SLE paths. As an extension of this result, we also compute the probability that an SLE$_\kappa$ path (with $0 < \kappa \le 4$) and a Brownian motion excursion do not intersect.
\end{abstract}

\section{Introduction}

\begin{quote}
``Though one can argue whether the scaling limits of interfaces in the Ising model are of
physical relevance, their identification opens possibility for computation of correlation functions and other objects of interest in physics.''\\
\phantom{x} \hfill \textsc{S.~Smirnov} (2007)
\end{quote}
The Schramm-Loewner evolution (SLE) is a one-parameter family of random growth processes in two dimensions introduced by O.~Schramm~\cite{schramm} while considering possible scaling limits of loop-erased random walk.   In the past several years, SLE techniques have been successfully applied to analyze a variety of two-dimensional statistical mechanics models including percolation, the Ising model, the $Q$-state Potts model, uniform spanning trees, loop-erased random walk, and self-avoiding walk.
Furthermore, SLE has provided a  mathematically rigorous framework for establishing various  predictions made by two-dimensional conformal field theory (CFT), and much current research is being done to further strengthen and explain the links between SLE and CFT; see, for example, \cite{dub, sheffdup, WernFried,kalle}.

In 2002, L.-P. Arguin and Y.~Saint-Aubin~\cite{saint} examined non-local observables in the 2d critical Ising model and using only techniques from conformal filed theory, they derived expressions for such things as the crossing probability of Ising clusters and contours intersecting the boundary of a cylinder. In particular, no mention of SLE was made in that work. In 2005, also using techniques exclusive to conformal field theory,  D.~Bauer, D.~Bernard, and K.~Kyt\"ol\"a~\cite{BBK} studied multiple Schramm-Loewner evolutions and statistical mechanics martingales. One consequence of their investigation was the computation of \emph{arch probabilities} (using their language) for the critical Ising model.

The primary purpose of the present work is to explain how the results of Arguin and Saint-Aubin~\cite{saint} as well as Bauer, Bernard, and Kyt\"ol\"a~\cite{BBK} for the Ising model can be derived in a mathematically rigorous manner by combining a recent result of S.~Smirnov~\cite{smirnovICM} with  the configurational measure on multiple SLE paths introduced by M.~Kozdron and G.~Lawler~\cite{KL}.  As an extension of this result, we also calculate the probability that an SLE$_\kappa$ path (with $0< \kappa \le 4$) and a Brownian excursion do not intersect.

\subsection{Towards a possible definition of a partition function for SLE} 

 In the case of a statistical mechanics lattice model, there are only a finite number of possible configurations. (Although this number is enormous, it is still finite.) Therefore, if a particular configuration $\omega'$ is given weight $\exp\{-H(\omega')/T\}$ where $T$ is the temperature and $H$ is the Hamiltonian, the probability of observing $\omega'$ is
 \begin{equation}\label{discrete}
 \Prob\{\omega'\} = \frac{\exp\{-H(\omega')/T\}}{
  \sum_{\omega}\exp\{-H(\omega)/T\}} = \frac{\exp\{-H(\omega')/T\}}{Z(T)}.
  \end{equation}
The normalizing factor $Z(T)$ is called the \emph{partition function} and it is well-known that this 
quantity encodes the statistical properties of a system in thermodynamic equilibrium.

However, in the scaling limit as the lattice spacing shrinks to 0, the ``number'' of configurations becomes infinite. 
From a physical point-of-view, when working with an ``infinite'' system one needs  an ``infinite'' term to be factored out so that the result is finite. The infinite factor, however, needs to be independent from the temperature, the shape of the domain, and other physically relevant quantities. Unfortunately, there is no consistent definition of partition function in physics and so the term is often used rather loosely, especially in the context of infinite systems.

As such, it is a challenge to mathematicians to make precise sense of what might be reasonably called a \emph{partition function for SLE}.  One way is to construct an object that possesses some of the characteristics of a partition function (in the physical sense). For instance, it might be chosen to satisfy a certain (physically relevant) differential equation. In the present manuscript we introduce an object that can, in this sense, be called a \emph{partition function for multiple SLE}. Mathematically, it is a normalizing factor that arises in the construction of a finite measure on multiple SLE paths, and satisfies the same differential equation as in Arguin and Saint-Aubin~\cite{saint}, as well as in Bauer, Bernard, and Kyt\"ol\"a~\cite{BBK}. (As we will indicate later on, there is some arbitrariness in the choice of normalization.)

It is worth noting that a treatment of partition functions has been recently proposed by J.~Dub\'edat~\cite{dub} that links SLE with the Euclidean free field by establishing identities between partition functions. A recent preprint by Lawler~\cite{LawNotes} explores another partition function view of SLE with some speculation about SLE in multiply connected domains.

\subsection{Outline}

The outline of the rest of this paper is as follows. In the next section we review the basics of SLE, and then in Section~\ref{SectConfig}, we review the configurational measure. We then review Smirnov's theorem for a single interface in the critical Ising model in Section~\ref{SectSmirnov}, and explain the theoretical predictions of Arguin and Saint-Aubin in Section~\ref{SectSaint}. In Section~\ref{SectPartition} we are able to construct the required partition function, and then show in Section~\ref{SectEquiv} how the  results of Arguin and Saint-Aubin~\cite{saint}, as well as Bauer, Bernard, and Kyt\"ol\"a~\cite{BBK},  can be recovered. Finally, in Section~\ref{SectSLEBMhit}, we extend the results of the previous sections with a theoretical result; namely, we compute the probability that an SLE$_\kappa$ path (with $0<\kappa \le 4$) and a Brownian excursion do not intersect.

\section{Review of SLE}\label{SectSLE}

It is assumed that the reader is familiar with the basics of SLE as described in any one of the general works for physicists such as \cite{BauBer6, C91, G1, Kag} or mathematicians such as~\cite{SLEbook, WernerNotes}. 
The purpose of this section is therefore to set notation we will use throughout and to review those properties of SLE germane for the present work.
Let $\C$ denote the set of complex numbers and write $\Half =\{z \in \C: \Im(z)>0\}$ to denote the upper half plane.
The \emph{chordal Schramm-Loewner evolution with parameter $\kappa > 0$ with the standard parametrization} (or simply SLE$_{\kappa}$)
is the random collection of conformal maps $\{g_t, \, t \ge 0\}$ of the upper half plane $\Half$ obtained by solving the initial value problem
\begin{equation}\label{SLEeqn}
\frac{\bd}{\bd t} \, g_t(z) = \frac{2}{g_t(z)-\sqrt{\kappa} W_t}, \;\;\; g_0(z)=z, 
\end{equation}
where $z \in \Half$ and $W_t$ is a standard one-dimensional Brownian motion with $W_0=0$. It is a hard theorem to prove that there exists a curve $\gamma:[0,\infty) \to \overline{\Half}$ with $\gamma(0)=0$ which generates the maps $\{g_t, \; t \ge 0\}$.
More precisely, for $z \in\Half$,  let $T_z$ denote the first time of explosion of the chordal Loewner equation~(\ref{SLEeqn}), and define the hull $K_t$ by
$K_t  = \overline{\{z \in \Half : T_z <t\}}$.
The hulls $\{K_t, \;t\ge 0\}$ are an increasing family of compact sets in $\overline{\Half}$ and $g_t$ is a conformal transformation of $\Half \setminus K_t$ onto $\Half$.  For all $\kappa>0$, there is a continuous curve $\{\gamma(t), \; t \ge 0\}$ with $\gamma:[0,\infty) \to \overline{\Half}$ and  $\gamma(0)=0$ such that $\Half \setminus K_t$ is the unbounded connected component of $\Half \setminus \gamma(0,t]$ a.s.  The behaviour of the curve $\gamma$ depends on the parameter $\kappa$.  If $0<\kappa \le 4$, then $\gamma$ is a simple curve with $\gamma(0,\infty) \subset \Half$ and $K_t = \gamma(0,t]$. If $4<\kappa <8$, then $\gamma$ is a non-self-crossing curve with self-intersections and $\gamma(0,\infty) \cap \R \neq \emptyset$. Finally, if $\kappa \ge 8$, then for this regime $\gamma$ is a space-filling, non-self-crossing curve.
Let $\mu^{\#}_{\Half}(0,\infty)$ denote the chordal SLE$_{\kappa}$ probability measure on paths in $\Half$ from 0 to $\infty$. Following Schramm's original definition~\cite{schramm}, 
if $D \subset \C$ is a simply connected domain and $z$, $w$ are distinct points in $\bd D$, then $\mu^{\#}_{D}(z,w)$, the chordal SLE$_{\kappa}$ probability measure on paths in $D$ from $z$ to $w$, is defined as the image of $\mu^{\#}_{\Half}(0,\infty)$ under a conformal transformation $f:\Half \to D$ with $f(0)=z$ and $f(\infty)=w$.

\begin{remark}
We are considering SLE$_{\kappa}$ as a measure on unparametrized curves. This means that it is sufficient to define SLE$_{\kappa}$ in $D$ from $z$ to $w$ to be the conformal image of SLE$_{\kappa}$ in $\Half$ from $0$ to $\infty$ under any conformal transformation with $0 \mapsto z$ and $\infty \mapsto w$. Of course, if $F: D \to \Half$ is a conformal transformation with $F(z)=0$ and $F(w)=\infty$, then $F$ is \emph{not} unique.  However, any other such transformation $\hat F$ must be of the form $\hat F = rF$ for some $r>0$. It is not too difficult to show that the definition of SLE$_{\kappa}$ in $D$ from $z$ to $w$ is then independent of the choice of transformation; see page~149 of~\cite{SLEbook}.
\end{remark}

As previously mentioned, a number of authors have been working to understand more fully the relationship between CFT and SLE. One form of this relationship comes in the interpretation of certain conformal field theory quantities in terms of $\kappa$, the variance parameter for the underlying Brownian motion driving process.
In particular, if we let
\begin{equation}\label{paramdefns}
    b = \frac{6 - \kappa}{2\kappa} \;\;\; {\rm and} \;\;\;  \ccharge = \frac{(\kappa-6)(8-3\kappa)}{2\kappa} = 1-\frac{3(\kappa-4)^2}{2\kappa}, 
\end{equation}
then $b$ is the \emph{boundary
scaling exponent} or \emph{boundary conformal weight} (also denoted $h_{1,2}$ in the CFT literature), and $\ccharge$ is the {\em central charge}.

\section{Review of the configurational measure}\label{SectConfig}

Early in the development of SLE, it was realized that interfaces of statistical mechanics models could be described in the scaling limit by a single chordal SLE path. Naturally, this led to the question of multiple interfaces and was the primary motivation for Bauer, Bernard, and Kyt\"ol\"a~\cite{BBK} to examine multiple SLE. More mathematical approaches were considered by Dub\'edat~\cite{dubeuler} who took a local, or infinitesmal, approach to the study of multiple SLE whereas Kozdron and Lawler~\cite{KL} viewed multiple SLE from a global, or configurational, point-of-view. The configurational approach, which we now recall, is  to view chordal SLE$_\kappa$ as not just a probability measure on paths connecting two specified points on the boundary, but
rather as a finite measure on paths that when normalized gives chordal SLE$_\kappa$ as defined by Schramm. This approach~\cite{KL} works in the case of simple paths, and so we restrict our consideration to SLE$_\kappa$ for $0<\kappa \leq 4$. For simplicity, the results are phrased in terms of the parameter $b$ (the boundary scaling exponent) which is related to $\kappa$ as in~(\ref{paramdefns}) by
\begin{equation*}   
b = \frac{6 - \kappa}{2\kappa} \;\;\; {\rm or}  \;\;\; \kappa = \frac{6}{2b+1}.
\end{equation*}
 Let $\mu^\#_{D,b,1}(z,w)$ denote the conformally invariant probability measure  
on chordal SLE$_{\kappa}$ paths from $z$ to $w$ in $D$ as defined in Section~\ref{SectSLE}. (Note that we wrote $\mu^\#_{D,b,1}(z,w)$ as $\mu^\#_{D}(z,w)$ in that section. We now want to emphasize the explicit dependence on $b$ and the fact that this is the measure on one path.) Define a kernel for the upper half plane $\Half$ by setting 
\begin{equation}\label{EPK}
H_{\Half, b,1}(0,\infty)=1 \;\;\; {\rm and} \;\;\;
              H_{\Half,b,1}(x,y) ={|y-x|^{-2b}} 
\end{equation}
for  $x$, $y \in \R = \bd \Half$. If $D$ is a simply connected domain with Jordan boundary and $z$, $w$ are distinct boundary points at which $\bd D$ is analytic, we now let $H_{D,b,1}(z,w)$ be determined by
\begin{equation}\label{eq1}%\label{hinvariance}
        H_{D,b,1}(z,w) =   |f'(z)|^b \, |f'(w)|^b
  \,  H_{f(D),b,1}(f(z),f(w))
\end{equation}
where $f:D \to f(D)$ is a conformal transformation. Finally, \emph{define} the SLE$_\kappa$ measure on
paths in $D$ from $z$ to $w$ by setting 
\begin{equation*}  
   Q_{D,b,1}(z,w) =  H_{D,b,1}(z,w) \; \mu^\#_{D,b,1}(z,w). 
\end{equation*}  
Note that this measure satisfies the conformal covariance rule
\begin{equation*}  
  f \circ Q_{D,b,1}(z,w)  =  |f'(z)|^b \, |f'(w)|^b
 \; Q_{f(D),b,1}(f(z),f(w))
 \end{equation*}  
which follows immediately from the conformal invariance
of  $\mu^\#_{D,b,1}(z,w)$ and the scaling rule~(\ref{eq1})
for $ H_{D,b,1}(x,y)$.

\begin{remark}
It is worth stressing that there is some arbitrariness possible in the definition of $H_{D,b,1}(z,w)$.  Motivated by conformal field theory, we want to define an object which satisfies the conformal covariance rule~(\ref{eq1}). Suppose that $D$ is a simply connected proper subset of $\C$ and $\bd D$ is locally analytic at $z$ and $w$. Suppose further that $D'$ is also a simply connected proper subset of $\C$ that is locally analytic at $z'$, $w' \in \bd D'$. It then follows that there exists a unique conformal transformation $f:D \to D'$ with $f(z)=z'$, $f(w)=w'$, and $|f'(w)|=1$. We call this the canonical transformation of $(D,z,w)$ onto $(D',z',w')$. In order to handle the case that $w=\infty$, we need to interpret things appropriately. We say that $\bd D$ is locally analytic at $w=\infty$ if $\bd h(D)$ is locally analytic at $0$ where $h(\zeta)=1/\zeta$. We interpret $|f'(w)|=1$ if $w=\infty$ (and $w' \neq \infty$) to mean that 
$|f(\zeta)-w| \sim |\zeta|^{-1}$ as $\zeta \to \infty$. Since a conformal transformation of the upper half plane $\Half$ onto itself with $\infty \mapsto \infty$ takes the form $f(z) = a_1z + a_2$ with $a_1, a_2 \in \R$ and $a_1>0$, 
in order to have $H_{\Half,b,1}(x,y) =   |f'(x)|^b \, |f'(y)|^b  \,  H_{\Half,b,1}(f(x),f(y))$
for $x$, $y \in \R$
 it must be the case that 
$H_{\Half,b,1}(x,y) =C{|y-x|^{-2b}}$
where $C>0$ is a constant.  If we now use the canonical transformation from $(\Half, 0, \infty)$ onto $(\Half, 0, 1)$ which is given by $f(z)=z/(1+z)$, then it follows that
$H_{\Half,b,1}(0,\infty) = |f'(0)|^{b} \, |f'(\infty)|^{b} \, H_{\Half,b,1}(0,1)  =C$.
We then, arbitrarily, choose $C=1$ so that $H_{\Half, b,1}(0,\infty)=1$
and 
$Q_{\Half,b,1}(0,\infty) =  \mu^\#_{\Half,b,1}(0,\infty)$ is the SLE probability measure on paths as originally defined by Schramm. This accounts for the declaration made in~(\ref{EPK}).
\end{remark}

We will now define the measures  $Q_{D,b,n} $ for positive
integers $n$.
As above, suppose that $D$ is a simply connected domain with Jordan boundary, and suppose that $z_1, \ldots, z_n, w_n, \ldots, w_1$ are $2n$ distinct points ordered counterclockwise on $\bd D$. Write $\z = (z_1,\ldots,z_n)$, $\w = (w_1,\ldots,                             
w_n)$, and assume that $\bd D$ is locally analytic at $\z$ and $\w$.  
Our goal is to define a measure on mutually avoiding $n$-tuples of simple paths $\bargamma = (\gamma^1, \ldots, \gamma^n)$ in $D$. More accurately, $\gamma^j$ is an equivalence class of curves such that there is a representation $\gamma^j:[0,1] \to \C$ which is simple and has $\gamma^j(0)=z_j$, $\gamma^j(1)=w_j$. Then
$Q_{D,b,n}(\z,\w)$,
 the $n$-path SLE$_\kappa$ measure in $D$, 
 is defined to be the measure
that  is  absolutely continuous with respect to
the product measure
\begin{equation*}  
Q_{D,b,1 }(z_1,w_1) \times \cdots\times Q_{D,b,1 }(z_n,w_n)
\end{equation*}  
with Radon-Nikodym derivative
$  Y({\bargamma}) =  Y_{D,b,\z,\w} (\gamma^1,\ldots,\gamma^n)$ given by
\begin{equation*}   %\label{nov21.1}
Y({\bargamma}) = 1\{ \gamma^k \cap \gamma^l
= \emptyset, \; 1 \leq k < l \leq n \} \;
\exp\left\{-\lambda \sum_{k=1}^{n-1}
             m(D;\gamma^k,\gamma^{k+1})\right\}
\end{equation*}  
where 
\begin{equation}\label{eqlambda}
\lambda = \frac{(6-\kappa)(8-3\kappa)}{4\kappa} = -\frac{\ccharge}{2}
\end{equation}
and  $m(D; V_1,V_2)$ denotes the Brownian loop measure of loops in $D$ that intersect both $V_1$ and $V_2$. For further details about the Brownian loop measure, consult~\cite{LawWloop}.
Finally, we define $H_{D,b,n}(\z,\w) = | Q_{D,b,n}(\z,\w)|$ to be the mass of the measure $Q_{D,b,n}(\z,\w)$, and note that it satisfies the conformal covariance property
\begin{equation*}  
H_{D,b,n} (\z,\w) = |f'(\z)|^b \, |f'(\w)|^b \,
     H_{f(D),b,n}(f(\z),f(\w))
\end{equation*}  
where we have written $f(\z) = (f(z_1), \ldots, f(z_n))$ and $f'(\z) = f'(z_1) \cdots f'(z_n)$. 
We end this section by summarizing the properties of the configurational measure. For proofs of the separate parts, see~Proposition~3.1, Proposition~3.2, Proposition~3.3 in~\cite{KL}.

\begin{theorem}[Properties of the configurational measure]
Suppose that $0 < \kappa \le 4$. Let $D$ be a simply connected domain with Jordan boundary, and let  $z_1, \ldots, z_n, w_n, \ldots, w_1$ be $2n$ distinct points ordered counterclockwise on $\bd D$. Write $\z = (z_1,\ldots,z_n)$, $\w = (w_1,\ldots,                             
w_n)$, and assume that $\bd D$ is locally analytic at $\z$ and $\w$.\\

\begin{itemize}%[000]
\item[\textbf{\emph{(a)}}] \emph{(Existence)}
For any $b \ge \frac{1}{4}$, the family of configurational measures $Q_{D,b,n}(\z,\w)$ as defined above is supported on $n$-tuples of mutually avoiding simple curves where each simple curve $\gamma^i$, $i=1,\ldots, n$, is chordal SLE$_{\kappa}$ from $z_i$ to $w_i$ 
with $\kappa = 6/(2b+1)$.\\

\item[\textbf{\emph{(b)}}] \emph{(Conformal Covariance)}
If $f :D \to f(D)$ is a conformal transformation, then
\begin{equation*}     
Q_{D,b,n} (\z,\w) = |f'(\z)|^b \, |f'(\w)|^b \,
     Q_{f(D),b,n}(f(\z),f(\w))
\end{equation*}  
where
$f(\z) = (f(z_1), \ldots, f(z_n))$ and $f'(\z) = f'(z_1) \cdots f'(z_n)$. \\

\item[\textbf{\emph{(c)}}] \emph{(Boundary Perturbation)}
Suppose $D \subset D' \subsetneq \C$ are simply
connected domains.    Then
$Q_{D,b,n}(\z,\w)$ is absolutely continuous with respect to
$Q_{D',b,n}(\z,\w)$ with Radon-Nikodym derivative equal to
\begin{equation*}  
\fl
\, Y_{D,D',b,n}(\z,\w)(\bargamma)
 =  1\{ \gamma^j \subset D, \;
j=1,\ldots,n\} 
\exp\{-\lambda \,m(D';\gamma^1 \cup \cdots \cup
  \gamma^n, D' \setminus D)\}
\end{equation*}  
where $m$ is the Brownian loop measure and $\lambda$ is given by~(\ref{eqlambda}).
  In particular, the Radon-Nikodym derivative is a conformal invariant.\\

\item[\textbf{\emph{(d)}}]  \emph{(Cascade Relation)}
For $1\leq j \leq n$, if
\begin{equation*}
\fl \, \z = (z_1,\ldots,z_n), \;\; \w = (w_1,\ldots,
w_n), \;\; \hatgamma = (\gamma^1,\ldots,\gamma^{j-1},
\gamma^{j+1},\ldots,\gamma^n),
\end{equation*}
\begin{equation*}
\fl \, \hat \z = (z_1,\ldots,z_{j-1}, z_{j+1},\ldots,
z_n) ,\;\; \hat \w = (w_1,\ldots,w_{j-1},w_{j+1},\ldots,w_n),
\end{equation*}
then the marginal measure on $\hatgamma$  in $Q_{D,b,n}(\z,\w)$
is absolutely continuous with respect to $Q_{D,b,n-1}(\hat \z,
\hat \w)$ with Radon-Nikodym derivative equal to
$H_{\hat D,b,1}(z_j,w_j)$. 
Here $\hat D$ is the subdomain of $D \setminus \hatgamma$ whose boundary includes
$z_j$, $w_j$. Moreover, the conditional distribution of $\gamma^j$ given
$\hatgamma$
is that of SLE$_\kappa$ from $z_j$ to $w_j$ in $\hat D$.
\end{itemize}

\end{theorem}

It is important to note that the construction just given is for a finite measure on $n$-tuples of mutually avoiding chordal SLE$_\kappa$ curves. The corresponding  probability measure is therefore given by
\begin{equation*}
\mu^\#_{D,b,n}(\z,\w) = \frac{Q_{D,b,n}(\z,\w)}{ H_{D,b,n}(\z,\w)}.
\end{equation*}

 \section{Smirnov's theorem for a single interface}\label{SectSmirnov}
 
Recent work by S.~Smirnov~\cite{smirnovICM} has established that the scaling limit of the  interface in the 2d Ising model at the critical temperature is SLE$_3$.  

To be precise, suppose that $D \subsetneq \C$ is a simply connected Jordan domain with distinct points $z$ and $w$ marked on the boundary. For every $N =1,2,3,\ldots$, let $(D_N,z_N, w_N)$ denote a simply connected, square lattice approximation to $(D,z,w)$,  and assume that $\{(D_N, z_N, w_N)\}$ converges in the Carath\'eodory sense as $N \to \infty$; see~\cite{KozALEA} for one way to construct such a sequence of discrete approximations to $(D,z,w)$.
Since the boundary of $D$ is a Jordan curve, the points $w$ and $z$ divide $\bd D$ into two arcs---the counterclockwise arc from $w$ to $z$ written $\bd^{+}$ and the counterclockwise arc from $z$ to $w$ written $\bd^{-}$. Let the corresponding subsets of $\bd D_N$ be denoted $\bd_N^{+}$ and $\bd_N^{-}$.
Now consider the Ising model at criticality on the lattice $(D_N, z_N, w_N)$ with boundary conditions of spin $+1$ at all points of $\bd_N^+$ and spin $-1$ at all points of $\bd_N^-$. (Without loss of generality, assume that both $z_N$ and  $w_N$ are $+1$.)
The result of Smirnov is that the discrete interface joining $z_N$ to $w_N$ and  separating $+1$ spins and $-1$ spins converges as $N \to \infty$ to a simple path whose law is given by the probability measure on chordal SLE$_3$ paths in $D$ from $z$ to $w$.  

\begin{remark}
Technically, Smirnov considers the Fortuin-Kastelyn random cluster representation of the Ising model on the square lattice. Introducing Dobrushin boundary conditions, namely wired on $\bd_N^+$ and dual-wired on $\bd_N^-$,  forces there to be a unique interface (on the medial lattice between the original lattice $D_N$ and its  dual-lattice) from $z_N$ to $w_N$ separating $+1$ spins and $-1$ spins; for details of the precise setup and statement, see~\cite{smirnovICM}.
\end{remark}

\section{Arguin and Saint-Aubin's theoretical predictions for two interfaces}\label{SectSaint}

\begin{center}
 \begin{figure}[h]
 \centering
\includegraphics{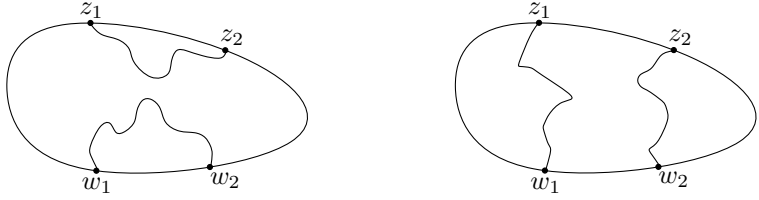}
 \caption{Configuration of Type~I (left) and Type~II (right).}\label{configtypesfig}
 \end{figure}
\end{center}

It also follows\footnote{No mathematical proof with all the details has been written down as of yet, although initial analysis suggests that it follows directly.} from Smirnov's work that if $w_1$, $w_2$, $z_2$, $z_1$ are four distinct marked boundary points labelled counterclockwise around $\bd D$, then the two interfaces of the Ising model at criticality with boundary changing operators at $w_{1,N}$, $w_{2,N}$, $z_{2,N}$, and $z_{1,N}$ converge as $N \to \infty$ to a pair of mutually avoiding simple paths whose law is that of a probability measure on pairs of mutually avoiding chordal SLE$_3$ paths. (This is explained more precisely in a remark in Section~\ref{SectPartition}.) There is, of course, the question of whether the multiple SLE$_3$ paths connect $w_1$ to $w_2$ and $z_1$ to $z_2$ or $z_1$ to $w_1$ and $z_2$ to $w_2$. Thus, using the language of Bauer, Bernard, and Kyt\"ol\"a~\cite{BBK}, there are two distinct \emph{arch-types} that may result. We prefer to use the phrase \emph{type of configuration} instead, and say that the resulting multiple interface configuration is of either Type~I if it joins  $w_1$ to $w_2$ and $z_1$ to $z_2$, or of Type~II if it joins $z_1$ to $w_1$ and $z_2$ to $w_2$; see Figure~\ref{configtypesfig}.

In the language of conformal field theory, an interface is a non-local observable, and Arguin and Saint-Aubin~\cite{saint}  used CFT techniques to give a prediction for the probability of a configuration of Type~II. They described the asymptotic behaviour of this probability using non-unitary representations that followed from the boundary scaling exponent $h_{1,2}$ of the Kac table.

Arguin and Saint-Aubin considered the Ising model at criticality on a half-infinite cylinder of radius 1. They represented  the half-infinite cylinder by the unit disk $\Disk$ and denoted by $\theta_j$, $j=1,..,4$, the four points along the boundary where the spin flips occurred. 

 They then conformally mapped the unit disk to the upper half plane, and argued that the four point correlation function of the field $\phi=\phi_{2,1}$ of conformal weight $\frac{1}{2}$ is 
\begin{equation*}
\langle \phi(z_1), \phi(w_1), \phi(z_2), \phi(w_2) \rangle 
 = \frac{1}{(z_1-w_1)(z_2-w_2)}  g\left( \frac{(z_1-w_1)(z_2-w_2)}{(z_1-z_2)(w_1-w_2)} \right)
 \end{equation*}
 where $g$ is a solution to the differential equation
\begin{equation}\label{sainteq}
3z(z-1)^2g''(z) + 2(z-1)(z+1)g'(z)-2zg(z)=0.
\end{equation}
This second order differential equation has two solutions---the first with exponent 0 and the second with exponent $\frac{5}{3}$. 
Arguin and Saint-Aubin argued that the solution with exponent $\frac{5}{3}$ corresponded to 
the probability of a configuration of Type~II, and then found
\begin{equation}\label{argsainteq1}
\fl \,\Prob\{{\rm config~of~Type~II}\}
= \frac{1}{2} - \frac{9}{20} \frac{\Gamma(\frac{1}{3})}{\Gamma(\frac{2}{3})^2 f^{(0)}(\xi)} \left( 
f^{(5/3)}(\xi) - \frac{\xi}{1-\xi} f^{(5/3)}(1-\xi)
\right)
\end{equation}
with
\begin{equation*}
f^{(0)}(\xi) = 1- \xi + \frac{\xi}{1-\xi} \;\;\; {\rm and} \;\;\; f^{(5/3)}(\xi) = 
 \frac{\xi^{5/3}}{1-\xi} F(-\case{1}{3}, \case{4}{3}, \case{8}{3}, \xi)
 \end{equation*}
where $F= {}_2F_1$ denotes the hypergeometric function
 and 
\begin{equation*}
\xi =  \frac{(z_1-w_1)(z_2-w_2)}{(z_1-z_2)(w_1-w_2)}
\end{equation*}
 denotes the \emph{cross-ratio}.
Furthermore, as $\xi \to 0$, it follows that
\begin{equation*}
\Prob\{{\rm config~of~Type~II}\} \sim 1 - \frac{10}{9} \frac{
\Gamma(\frac{2}{3})^2
}{\Gamma(\frac{1}{3})
}\xi^{5/3}  + \Or(\xi^2).
\end{equation*}
In Section~\ref{SectEquiv} we explain how to recover the result~(\ref{argsainteq1}) rigorously using SLE.
  
\section{Definition of a partition function for two paths and a crossing probability calculation}\label{SectPartition}

Recall from Section~\ref{SectConfig} that    $H_{D,b,n}(\z,\w)$ is defined to be the mass of the configurational measure $Q_{D,b,n}(\z,\w)$ and that 
  $H_{D,b,n}$ satisfies the  scaling rule
\begin{equation*}
 H_{D,b,n} (\z,\w) = |f'(\z)|^b \, |f'(\w)|^b \,
     H_{f(D),b,n}(f(\z),f(\w)) . 
\end{equation*}
If we now define 
\begin{equation}\label{corrfcn}
  \tilde H_{D,b, n}(\z,\w) = \frac{H_{D,b ,n} (\z,\w)}
     {H_{D,b,1}(z_1,w_1) \cdots H_{D,b,1}(z_n,w_n)},
\end{equation}
then   $\tilde H_{D,b, n}(\z,\w)$ is a conformal \emph{invariant}. Thus, by conformal invariance, it suffices to work in $D=\Half$.

In the case of two paths, an explicit calculation is possible and is given by the following proposition which has appeared in a number of places. It was first stated in a rigorous mathematical context by Dub\'edat~\cite{dubeuler} using an infinitesmal approach, and was derived using CFT by Bauer, Bernard, and Kyt\"{o}l\"{a}~\cite{BBK}. A detailed derivation first appeared in~\cite{KL}. As we will see shortly, the special case of the Ising model actually appeared earlier in Arguin and Saint-Aubin~\cite{saint}.

\begin{proposition} \label{dec1.prop1}
Consider the upper half plane $\Half$, and  let $0 < x < y <\infty$. If $b \geq \frac{1}{4}$, then
\begin{equation}  \label{nov19.2}
\tilde H_{\Half,b,2}((0,x),(\infty,y)) = \frac{\Gamma(2a) \, \Gamma(6a-1)}
            {\Gamma(4a) \, \Gamma(4a-1)} \, (x/y)^a \,
                F(2a,1-2a,4a; x/y)
\end{equation}
where $F={}_2F_1$ denotes the hypergeometric function
and $a = 2/\kappa = (2b+1)/3$.
\end{proposition}

The proof of this proposition in~\cite{KL} is accomplished by finding and then solving a differential equation satisfied by $\tilde H_{\Half,b,2}((0,x),(\infty,y))$. By scaling, we can write  $\tilde H_{\Half,b,2}((0,x),(\infty,y)) = \psi(x/y)$ for some function $\psi=\psi_{\Half,b}$ of one variable. We then show that the ODE satisfied by $\psi$ is
\begin{equation*}%\label{hypergeomFomin}
\fl \, u^2 \, (1-u)^2 \,\psi''(u)  +  2 \,
 u \, (a-u + (1-a) \, u^2) \, \psi'(u) -a(3a-1) (1-u)^2 \, \psi(u) =0
\end{equation*}
where $a=2/\kappa$. In the particular case that $\kappa=3$ so that $a=\frac{2}{3}$, this differential equation reduces to 
\begin{equation*}%\label{hypergeomFomin}
 3u^2 \, (1-u) \,\psi''(u)  +  2 \,
 u \, (2-u) \, \psi'(u) -2(1-u) \, \psi(u) =0.
\end{equation*}
If we change variables by setting $g(z) = \psi(1-z)$, then $g$ satisfies
\begin{equation*}%\label{hypergeomFomin}
 3z(z-1)^2 \, g''(z)  + 2 \, (z-1) \, (z+1) \, g'(z) -2z \, g(z) =0
\end{equation*}
which is exactly~(\ref{sainteq}) above.

\begin{remark}
It is important to note that the restriction to $b \ge \frac{1}{4}$ is needed to guarantee that $0 < \kappa \le 4$. Formally, if we plug in $\kappa=6$, then  we recover Cardy's formula for percolation; however, constructing a configurational measure on non-crossing SLE paths in the non-simple regime ($4<\kappa<8$) is still an open problem. 
\end{remark}

We now explain how Proposition~\ref{dec1.prop1} can be used to calculate a crossing probability for two SLE$_\kappa$ paths ($0 <\kappa \le 4$). Choosing $\kappa=3$ as a special case yields the desired result of Arguin and Saint-Aubin~\cite{saint}, and of Bauer, Bernard, and Kyt\"ol\"a~\cite{BBK}, for the critical Ising model. By conformal invariance, it is enough to work in the upper half plane $\Half$ with boundary points $0$, $1$, $\infty$, and $x$ where $0<x<1$ is a real number.  
 The possible interface configurations are therefore of two types, namely (I) a simple curve connecting $0$ to $\infty$ and a simple curve connecting $x$ to $1$, or (II) a simple curve connecting $0$ to $x$ and a simple curve connecting $1$ to $\infty$.  
The configurational measure corresponding to Type~I is 
 \begin{equation*}Q_{\Half,b,2}((0,x), (\infty, 1))\end{equation*}
 and the configurational measure corresponding to Type~II is
 \begin{equation*}Q_{\Half,b,2}((x,1), (0,\infty)).\end{equation*}
By the symmetry of chordal SLE about the imaginary axis, however, 
 \begin{equation*}Q_{\Half,b,2}((x,1), (0,\infty)) = Q_{\Half,b,2}((0,1-x), (\infty,1)).\end{equation*}
The partition function corresponding to a configuration of Type~I is (defined as)
 \begin{equation*}Z_{b,I}(x) := H_{\Half,b,2}((0,x), (\infty,1))\end{equation*}
 and the partition function corresponding to a configuration of Type~II is (defined as)
  \begin{equation*}Z_{b,II}(x) := H_{\Half,b,2}((0,1-x), (\infty,1)) = Z_{b,I}(1-x).\end{equation*}
Therefore, the probabilities of a configuration  of Type~I and of a configuration of Type~II
are given by
\begin{equation}\label{configprobs}
\frac{Z_{b,I}(x)}{Z_{b,I}(x) + Z_{b,II}(x)}
\;\;\; {\rm and} \;\;\;
\frac{Z_{b,II}(x)}{Z_{b,I}(x) + Z_{b,II}(x)} =
\frac{Z_{b,I}(1-x)}{Z_{b,I}(x) + Z_{b,II}(x)},
\end{equation}
respectively.

\begin{remark}
As indicated in Section~\ref{SectConfig}, we chose to normalize our kernel in such a way that $H_{\Half,b,1}(0,\infty)=1$. Thus, there is no arbitrary constant in our definition of either $Z_{b,I}(x)$ or  $Z_{b,II}(x)$. Suppose, however,  that we had normalized our kernel differently, say  $H_{\Half,b,1}(0,\infty)=C$ for some $C>0$. Although both $Z_{b,I}(x)$ and $Z_{b,II}(x)$ would now depend on $C$,  the ratios in~(\ref{configprobs}) would not.
\end{remark}

\begin{remark}
To be precise, the construction in Section~\ref{SectConfig} only defines the configurational measure for a given type of configuration.  If we want to consider configurations without regard to type, then we need to define a measure supported on mutually-avoiding pairs of curves of either type. Of course, such a measure is given by the sum of the configurational measures of Types~I and~II, repectively. The mass of this measure is $Z_{b,I}(x) + Z_{b,II}(x)$, and so the probability measure on mutually-avoiding pairs of curves of either type is
\begin{equation*}
\Prob = \frac{Q_{\Half,b,2}((0,x), (\infty, 1)) + Q_{\Half,b,2}((x,1), (0,\infty))}{Z_{b,I}(x) + Z_{b,II}(x)}.
\end{equation*}
Thus, if $A$ is the event $A = \{{\rm config~of~Type~I}\}$, then
\begin{eqnarray*}
\Prob(A) &= \frac{Q_{\Half,b,2}((0,x), (\infty, 1))(A) + Q_{\Half,b,2}((x,1), (0,\infty))(A)}{Z_{b,I}(x) + Z_{b,II}(x)}\\
&= \frac{Z_{b,I}(x) + 0}{Z_{b,I}(x) + Z_{b,II}(x)}\\
\end{eqnarray*}
and, similarly, for $\Prob(A^c) = \Prob\{{\rm config~of~Type~II}\}$ as in~(\ref{configprobs}).
We can now give a more careful statement of the consequence of Smirnov's work mentioned at the beginning of Section~\ref{SectSaint}, namely that if $\Prob_N$ denotes the probability measure for the two interfaces on the $1/N$-scale lattice, then $\Prob_N$ converges weakly to $\Prob$.
\end{remark}

Now by~(\ref{EPK}) and~(\ref{corrfcn}), we know that
\begin{eqnarray*}
H_{\Half,b,2}((0,x),(\infty,1)) &=H_{\Half,b,1}(0,\infty) \cdot
H_{\Half,b,1}(x,1) \cdot \tilde  H_{\Half,b,2}((0,x),(\infty,1))\\
&= (1-x)^{-2b} \tilde H_{\Half,b,2}((0,x),(\infty,1)) 
\end{eqnarray*}
so that Proposition~\ref{dec1.prop1} yields
%\begin{eqnarray*}
\begin{equation*}
\fl \, Z_{b,I}(x)=H_{\Half,b,2}((0,x),(\infty,1))
 = \frac{\Gamma(2a) \, \Gamma(6a-1)}  {\Gamma(4a) \, \Gamma(4a-1)} \, x^a (1-x)^{-2b}\, F(2a,1-2a,4a; x).
\end{equation*}
%\end{eqnarray*}
Using~(15.3.3) of~\cite{AbSteg}, we can write
\begin{equation}\label{absteg}
F(2a,1-2a,4a; x) = (1-x)^{4a-1}F(2a, 6a-1,4a;x).
\end{equation}
If we also note that $a=2/\kappa$ so that~(\ref{paramdefns}) implies $2b = (6-\kappa)/\kappa = 3a-1$, then we can write
\begin{equation*}Z_{b,I}(x)= \frac{\Gamma(2a) \, \Gamma(6a-1)}  {\Gamma(4a) \, \Gamma(4a-1)} \, x^a (1-x)^{a}\, F(2a, 6a-1,4a;x)\end{equation*}
and
\begin{equation*}Z_{b,II}(x)= \frac{\Gamma(2a) \, \Gamma(6a-1)}  {\Gamma(4a) \, \Gamma(4a-1)} \, x^a (1-x)^{a}\, F(2a, 6a-1,4a;1-x).\end{equation*}
Hence, we conclude from~(\ref{configprobs}) that
\begin{equation}\label{genform}
\fl \, \Prob\{{\rm config~of~Type~I}\}
%&=\frac{Z_{b,I}(x)}{Z_{b,I}(x) + Z_{b,II}(x)}\nonumber\\
=\frac{F(2a, 6a-1,4a;x)}{F(2a, 6a-1,4a;x)+ F(2a, 6a-1,4a;1-x)}
\end{equation}
and
\begin{equation}\label{genformII}
\fl \, \Prob\{{\rm config~of~Type~II}\}
%&=\frac{Z_{b,II}(x)}{Z_{b,I}(x) + Z_{b,II}(x)}\nonumber\\
=\frac{F(2a, 6a-1,4a;1-x)}{F(2a, 6a-1,4a;x)+ F(2a, 6a-1,4a;1-x)}.
\end{equation}

\section{Summary of results for the 2d critical Ising model}\label{SectEquiv}

\begin{center}
\begin{figure}[h]
\centering
\includegraphics[height=1.5in]{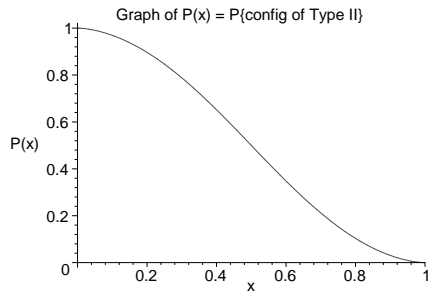}
\caption{Graph of $P(x)=P_1(x)=P_2(x)=P_3(x)$ for $0\le x\le 1$.}\label{graphfig}
\end{figure}
\end{center}

In the particular case of the 2d critical Ising model (in which case $\kappa=3$), then~(\ref{genformII}) yields the probability of a configuration of Type~II as follows:
\begin{equation*}P_1(x) = \frac{F(\frac{4}{3}, 3,\frac{8}{3};1-x)}{F(\frac{4}{3}, 3,\frac{8}{3};x)+ F(\frac{4}{3}, 3,\frac{8}{3};1-x)}.\end{equation*}
Using~(15.3.3) of~\cite{AbSteg} (as in~(\ref{absteg}) above) 
Arguin and Saint-Aubin~\cite{saint} express the same probability~(\ref{argsainteq1})  as
\begin{equation*}P_2(x) = \frac{1}{2} - \frac{9}{20} \frac{\Gamma(\frac{1}{3})}{\Gamma(\frac{2}{3})^2} \left[\frac{x^{5/3}(1-x)^{5/3}}{1-x+x^2}\right] \left[ F(\case{4}{3}, 3,\case{8}{3};x)- F(\case{4}{3}, 3,\case{8}{3};1-x) \right]\end{equation*}
whereas it is given in Bauer, Bernard, and Kyt\"ol\"a~\cite{BBK} as
\begin{equation*}P_3(x) =\left( \int_0^1 \frac{y^{2/3}(1-y)^{2/3} }{(1-y+y^2)^2} \, \rmd y
\right)^{-1} \int_x^1 \frac{y^{2/3}(1-y)^{2/3} }{(1-y+y^2)^2} \, \rmd y 
.\end{equation*}
It is not at all obvious that these three expressions are identical.  However, since all three represent the same physical observable (and since each was obtained by solving the same differential equation), it \emph{must} be the case that $P_1(x)=P_2(x)=P_3(x)$ for $0 \le x \le 1$; see Figure~\ref{graphfig}.

The easiest way to verify their equivalence is simply to check directly that each satisfies the required differential equation with the given boundary conditions. It is also possible to  verify algebraically that these three expressions are the same by converting all of the hypergeometric functions into associated Legendre functions of the first kind.

\begin{remark}
The calculation of $P_1(x)$ follows from SLE-techniques in a mathematically rigorous way, and it provides an explanation for the results of
Arguin and Saint-Aubin as well as Bauer, Bernard, and Kyt\"ol\"a. The key point is that the result of Smirnov tells us precisely what is meant by a scaling limit of the Ising model, namely  the interface separating $+1$ spins from $-1$ spins viewed as a probability measure on curves converges weakly to the law of choral SLE$_3$. Thus, by choosing $\kappa=3$ we should be able to use SLE to recover results from CFT for the Ising model such as the one that  Arguin and Saint-Aubin derived.
% We are using Smirnov's result simply to tell us that $\kappa=3$ corresponds to the scaling limit of interfaces in the critical Ising model.  Unfortunately, Smirnov's result does not enable us to immediately conclude that the scaling limit of the probability of a configuration of Type~I (or Type~II) is given by~(\ref{configprobs}). 
\end{remark}

\section{Intersection probabilities for SLE$_\kappa$, $0 < \kappa \le 4$, and a Brownian excursion}\label{SectSLEBMhit}

\begin{figure}[h]
 \centering
 \includegraphics[height=1.5in]{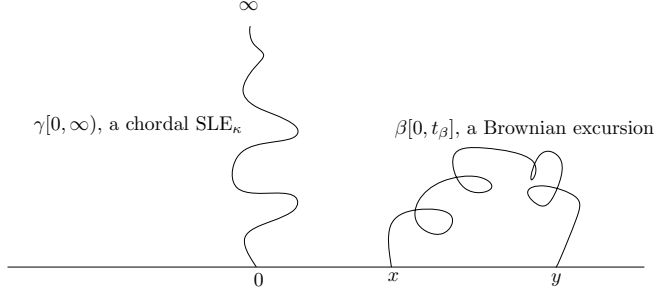}
\caption{Schematic representation of $\Prob\{ \gamma[0,\infty) \cap \beta[0,t_{\beta}] = \emptyset \}$.}\label{SLEfig}
\end{figure}

The techniques that were used in~\cite{KL} to derive Proposition~\ref{dec1.prop1} leads to a calculation of the probability that an SLE$_2$ path and a Brownian excursion do not intersect. This was the key in establishing the scaling limit of Fomin's identity for loop-erased random walk~\cite{KozCR}. In this section, we extend those ideas to compute the probability that an SLE$_\kappa$ path and a Brownian excursion do not intersect. This event is illustrated in Figure~\ref{SLEfig}. 

\begin{theorem}\label{FI_SLE}
Suppose that $0<x<y<\infty$ are real numbers and let $\beta:[0,t_{\beta}] \to \overline{\Half}$ be a Brownian excursion from $x$ to $y$ in $\Half$. If $\gamma:[0,\infty) \to \overline{\Half}$ is a chordal SLE$_{\kappa}$ from $0$ to $\infty$ in $\Half$, then 
\begin{eqnarray}\label{scalefomin}
\fl \Prob\{ \,\gamma[0,\infty) \cap \beta[0,t_{\beta}] = \emptyset\, \}
=\frac{\Gamma(2a)\Gamma(4a+1)}{\Gamma(2a+2)\Gamma(4a-1)} \, (x/y)\,  F(2, 1-2a, 2a+2; x/y)
\end{eqnarray}
where $F= {}_2F_1$ is the hypergeometric function and $a=2/\kappa$.
\end{theorem}
 
Since the proof of this theorem is similar to the proof of Theorem~6.1 in~\cite{KozCR}, we omit many details.

\begin{proof}
Let
$\intprob(x,y) = \Prob\{ \gamma[0,\infty) \cap \beta[0,t_{\beta}] = \emptyset  \}$. Using It\^o's formula, it can be shown that $\intprob$ satisfies the differential equation
\begin{equation}\label{ItoEqn}
  -a\left(\frac 1x - \frac 1y \right)^2 \, \intprob  + \frac{a}{x} \, 
    \frac{\bd \intprob}{\bd x}   
+ \frac{a}{y} \, \frac{\bd \intprob}{\bd y} + \frac 12 \, \frac{\bd^2 \intprob}{\bd x^2} 
     + \frac 12 \, \frac{\bd^2 \intprob}{\bd y^2}
   + \frac{\bd^2 \intprob}{\bd x\bd y}=0.
\end{equation}
SLE scaling implies that the probability in question only depends on the ratio $x/y$, and so
$\intprob(x,y) = \varphi(x/y)$ for some function $\varphi = \varphi_{\Half,b}$ of one variable. Thus, we find
\begin{equation*}  
\fl \, \frac{\bd \intprob}{\bd x} = y^{-1} \varphi'(x/y), \;\;\; \frac{\bd \intprob}{\bd y} = - xy^{-2} \varphi'(x/y), \;\;\;
   \frac{\bd^2 \intprob}{\bd x^2}= y^{-2}  \varphi''(x/y), 
\end{equation*}
\begin{equation*} 
\fl \,
 \frac{\bd^2 \intprob}{\bd y^2} = 2 x y^{-3} \varphi'(x/y) + x^2 y^{-4}   \varphi''(x/y), \;\;\;
 \frac{\bd^2 \intprob}{\bd x\bd y}= - y^{-2}  \varphi'(x/y) - x y^{-3}\varphi''(x/y),\end{equation*}
so that after substituting into~(\ref{ItoEqn}), multiplying by $y^2$, letting $u=x/y$,
and combining terms, we have
%\begin{equation*}%\begin{equation*}
% u^2  (1-u)^2\varphi''(u)  +    2 
% u  (a-u+(1-a)u^2)  \varphi'(u) -2a (1-u)^2  \varphi(u) =0
%\end{equation*}%\end{equation*}
%or, equivalently, 
\begin{equation}\label{hypergeomFominXX}
u^2 (1-u)\varphi''(u)  +    2  u  (a +(a-1)u)  \varphi'(u) -2a (1-u)  \varphi(u) =0
\end{equation}
using the constraint $0<u<1$.
The second-order ordinary differential equation~(\ref{hypergeomFominXX}) has regular singular points at $0$, $1$, and $\infty$, and so we know that it is possible to 
transform it into a hypergeometric differential equation.
By writing~(\ref{hypergeomFominXX}) as 
\begin{equation}\label{hypergeomFomin2-pform}
\varphi''(u)  + \left[ \frac{2a}{u} + \frac{2-4a}{u-1} \right] \varphi'(u) + \left[ \frac{2a}{u} - 2a \right]  \frac{\varphi(u)}{u(u-1)}  =0
\end{equation}
we see that we have a case of Riemann's differential equation whose complete set of solutions (see~(15.6.1) and~(15.6.3) of~\cite{AbSteg}) can be denoted by Riemann's $P$-function
\begin{equation*}\varphi(u) = P 
\left\{
\begin{array}{cccc}
0 &\infty &1 &\\
1 &-2a &4a-1 &u\\
-2a &1 &0&\\
\end{array}
\right\}.\end{equation*}
By now considering~(15.6.11) of~\cite{AbSteg}, the transformation formula for Riemann's $P$-function for reduction to the hypergeometric function, we see that the appropriate change-of-variables to apply is  $\psi(u) = u^{-1} (1-u)^{1-4a} \varphi(u)$ noting that this is permitted by the constraint $0<u<1$.
 Thus,~(\ref{hypergeomFominXX}) implies
\begin{equation}\label{hypergeomFomin3XX}
 u  (1-u) \psi''(u)  +  (2a+2-(6a+2)u) \psi'(u) -2a(4a+1) \psi(u) =0.
\end{equation}
We see that~(\ref{hypergeomFomin3XX}) is now a well-known hypergeometric differential equation~\cite{AbSteg} whose general solution is given by
\begin{equation*}
\psi(u)= C_1 F (2a,  4a+1, 2a+2; u) +C_2 u^{-1-2a}F(-1, 2a, -2a; u).
\end{equation*}
and so
\begin{equation*}
\fl \, \varphi(u) =  u(1-u)^{4a-1} \left[
 C_1 F(2a, 4a+1, 2a+2; u)
+C_2 u^{-1-2a}  F(-1, 2a, -2a; u)\right].
\end{equation*}
Using equation~(15.3.3) of~\cite{AbSteg} we find
\begin{equation*}F (2a, 4a+1, 2a+2; u) = (1-u)^{1-4a}F (2, 1-2a, 2a+2; u)\end{equation*}
which implies that
\begin{equation*}
\varphi(u)=
 C_1u \, F (2, 1-2a, 2a+2; u)+C_2 u^{-2a}(u-1)^{4a-1} F(-1, 2a, -2a; u).
\end{equation*}
However, it follows immediately from the continuity of the Brownian excursion measure~\cite{KozALEA} and the fact that $\gamma(0,\infty) \cap \R = \emptyset$ when $0 < \kappa \le 4$ that $\varphi(u) \to 0$ as $u \to 0+$ 
and $\varphi(u) \to 1$ as $u \to 1-$.
This implies $C_2=0$ and 
\begin{equation*}
\fl \, C_1^{-1}=F (2, 1-2a, 2a+2; 1) = \lim_{u \to 1-} F (2, 1-2a, 2a+2; u) =  
\frac{\Gamma(2a+2)\Gamma(4a-1)}
{\Gamma(2a)\Gamma(4a+1)}.
\end{equation*}
Thus, 
\begin{equation*}\varphi(u)= \frac{\Gamma(2a)\Gamma(4a+1)}{\Gamma(2a+2)\Gamma(4a-1)} \, u \, F (2, 1-2a, 2a+2; u)\end{equation*} 
and so~(\ref{scalefomin}) follows as required. 
\end{proof}      

\begin{remark}
Theorem~\ref{FI_SLE} provides another example of an SLE observable. Hence, the primary application of Theorem~\ref{FI_SLE} to a physical situation seems to be as a way to provide evidence that a particular statistical mechanics lattice model interface has an SLE limit. A conjectured value of $\kappa$ may be found, or verified, by approximating the probability that a Brownian excursion and an interface intersect,  and then comparing the result to that given in this theorem. In order to actually do this numerically, however, there are a number of issues with which one must contend. These include selecting a lattice with which to work, defining and then simulating an appropriate interface, and then simulating a simple random walk excursion on the lattice (since simple random walk excursions converge to Brownian excursions; see~\cite{KozALEA}, for instance). 

As an example where the observable from this theorem might be applied, consider the recent work of Bernard, Le Doussal, and Middleton~\cite{BDM}. They perform several statistical tests of the hypothesis that zero-temperature Ising spin glass domain walls are described by an SLE$_\kappa$, and working on the triangular lattice they find numerically these domain walls to be consistent with $\kappa =2.32 \pm 0.08$. 
Among the  observables studied in~\cite{BDM} that led to this conclusion is the SLE left-passage probability (which, incidentally, is also given in terms of a hypergeometric function). The work of Bernard, Le Doussal, and Middleton extends earlier work of 
Amoruso, Hartmann, Hastings, and Moore~\cite{AHHM} who presented numerical evidence that the techniques of CFT might be applicable to two-dimensional Ising spin glasses, and that such domain walls might be described by a suitable SLE. In particular, the observable studied in~\cite{AHHM} was the fractal dimension of the domain walls. 

The transition probabilities for simple random walk excursions on the triangular lattice can be computed. This means that such random walks can be simulated, and so it seems possible that the numerical techniques used in either~\cite{BDM} or~\cite{AHHM} could actually be applied for the observable of Theorem~\ref{FI_SLE}.
\end{remark}

\section{Conclusion}

The construction of the configurational measure on $n$-tuples of mutually avoiding, simple SLE paths  by Kozdron and Lawler~\cite{KL} leads to a possible definition of a partition function for SLE. Using this definition, a mathematically rigorous proof can be given for certain theoretical predictions about the 2d critical Ising model that Arguin and Saint-Aubin~\cite{saint} originally derived using only CFT techniques (i.e., no SLE mentioned in their work).  As well, this gives  a mathematically rigorous derivation of the general results of Bauer, Bernard, and Kyt\"ol\"a~\cite{BBK} concerning crossing probabilities for two interfaces in the simple ($0<\kappa \le 4$) regime that they derived previously using CFT techniques. It also leads to the calculation of the probability that an SLE$_\kappa$ path (with $0<\kappa \le 4$) and a Brownian excursion do not intersect.

\section*{Acknowledgements}

This paper had its origin at the \emph{Workshop on Stochastic Loewner Evolution and Scaling Limits} held in August 2008 at the Centre de Recherches Math\'ematiques. Thanks are owed to Yvan Saint-Aubin who brought the work~\cite{saint} to the attention of the author, as well as to the Natural Sciences and Engineering Research Council (NSERC) for providing financial support through the Discovery Grant program. 

%\section*{References}

\end{document}